\documentclass{ws-procs975x65}            
\begin{document}
\title{Quantum energy inequalities in integrable quantum field theories}

\author{D.~Cadamuro}

\address{Mathematisches Institut,
Georg-August Universit\"at G\"ottingen,\\
Bunsenstra\ss{}e 3-5,
D-37073 G\"ottingen, 
Germany\\
E-mail: daniela.cadamuro@mathematik.uni-goettingen.de}

\begin{abstract}
In a large class of factorizing scattering models, we construct candidates for the local energy density on the one-particle level starting from first principles, namely from the abstract properties of the energy density. We find that the form of the energy density at one-particle level can be fixed up to a polynomial function of energy. On the level of one-particle states, we also prove the existence of lower bounds for local averages of the energy density, and show that such inequalities can fix the form of the energy density uniquely in certain models.
\end{abstract}


\bodymatter

\section{Introduction}

The role and properties of the energy density are fundamental in quantum field theory.  On curved backgrounds the energy density links to the geometry of the spacetime via the Einstein field equations. In Minkowski space, it represents a local observable of particular physical significance, while in 2d conformal field theory the algebra of the stress-energy tensor is completely fixed up to a scalar (central charge), and  the existence of this large symmetry constrains significantly the structure of the theory.

In typical models of \emph{classical} field theory the energy density is positive, e.g., in electrodynamics. This has interesting consequences in general relativity, where it implies certain restrictions on the geometry of spacetime, excluding ``exotic'' configurations such as wormholes, warp drives and time machines. (See, e.g., the Penrose and Hawking singularity theorems\cite{HawPen1970}, positive mass theorems and Hawking's chronology protection results\cite{Hawking:1992}.)

In quantum field theory, even in flat spacetime (not only on curved background) the energy density can be negative, while the Hamiltonian $H$ is still nonnegative. However, one expects that the smeared energy density, $T^{00}(g^2) = \int g^2(t) T^{00}(t,0)$, with some fixed smooth real-valued test function $g$, cannot be indefinitely negative:\cite{Ford:1991} Certain lower bounds hold, the so called \emph{quantum energy inequalities} (QEIs). A \emph{state-independent} QEI takes the form
\begin{equation}\label{stinqei}
\langle \varphi, T^{00}(g^2) \varphi \rangle \geq -c_g \|\varphi  \|^2
\end{equation}
for all (suitably regular) state vectors $\varphi$, where $c_g > 0$ is a constant depending on the test function $g$. 

We note that in some theories, e.g., the non-minimally coupled scalar field in a curved spacetime, only a weaker form of this inequality can hold, a so called \emph{state-dependent} QEI, where the right hand side of Eq.~\eqref{stinqei} depends on the total energy of the state $\varphi$.

State-independent QEIs have been proved for the linear scalar field, linear Dirac field, linear vector field (both on flat and curved spacetime), the Rarita-Schwinger field, and for 1+1d conformal fields (see Ref.~\citenum{Fewster:lecturenotes} for a review). State-dependent QEIs were established in a model-independent setting\cite{BostelmannFewster:2009}  for certain ``classically positive'' expressions, but their relation to the energy density is unclear. 

Only recently QEIs have been obtained in the case of \emph{self-interacting} quantum field theories, the first example being the massive Ising model\cite{BostelmannCadamuroFewster:ising}. This model is an example of a specific class of self-interacting theories on 1+1 dimensional Minkowski space, so called \emph{quantum integrable models}\cite{BabujianFoersterKarowski:2006}. Other examples include the sinh-Gordon, the sine-Gordon and the nonlinear $O(N)$-invariant $\sigma$-models.

There has been recent interest into these models from the side of rigorous quantum field theory. In particular, a large class of these theories were constructed from a prescribed factorizing S-matrix using operator-algebraic techniques\cite{Lechner:2008}. They describe interacting relativistic particles and are characterized by infinitely many conserved currents, implying that the particle number is preserved during the scattering process and that the full scattering matrix is completely determined by the two-particle scattering function, hence called \emph{factorizing}. 

Here we consider such models with one species of massive scalar bosons and without bound states.  In this large class, we are interested in the stress-energy tensor $T^{\alpha \beta}$ evaluated in \emph{one-particle states}. At that level, we investigate the existence of (state-independent) QEIs, but also the uniqueness of $T^{\alpha \beta}$ itself, the existence of states with negative energy density, and the lowest eigenvalue in the spectrum of $T^{00}(g^2)$.

\section{Stress-energy tensor in one-particle states}

In models derived from a classical Lagrangian, such as the sinh-Gordon model (see Ref.~\citenum{FringMussardoSimonetti:1993}), a candidate for the energy density can be computed directly from the Lagrangian. However, there are examples of integrable models which are not associated with a Lagrangian (e.g. the generalized sinh-Gordon model in Table 1 of Ref.~\citenum{BostelmannCadamuro:oneparticle}). 

It is therefore useful to obtain an intrinsic characterization of the stress energy tensor $T^{\alpha \beta}$ at the \emph{one-particle level} starting from ``first principles'', namely from the generic properties of this operator. We write $T^{\alpha \beta}$ at one-particle level as an integral kernel operator:
\begin{equation}
\langle \varphi, T^{\alpha \beta}(g^2) \psi \rangle = \int \overline{\varphi(\theta)}F^{\alpha \beta}(\theta,\eta)\psi(\eta),
\end{equation}
where $\varphi, \psi$ are vectors in the single-particle space $L^2(\mathbb{R})$. 
There are various restrictions on the form of the kernel $F^{\alpha \beta}$. The fact that the energy density is a \emph{local} observable implies that $F^{\alpha \beta}$ fulfills certain analyticity, symmetry and boundedness properties\cite{BostelmannCadamuro:characterization}. Additional conditions come from the specific properties of the stress-energy tensor, namely tensor symmetry, covariance under Poincar\'e transformations and spacetime reflections, the continuity equation ($\partial_{\alpha}T^{\alpha \beta}=0$), and the fact that the $(0,0)$-component of the tensor integrates to the Hamiltonian ($\int dx\; T^{00}(t,x) = H$).

Starting from these general properties, we can determine $T^{\alpha \beta}$ in one-particle matrix elements up to a certain polynomial factor. More specifically, we find (see Proposition~3.1 of Ref.~\citenum{BostelmannCadamuro:oneparticle}) that functions $F^{\alpha \beta}$ are compatible with the requirements above if, and only if, there exists a real polynomial $P$ with $P(1)=1$ such that
\begin{equation}
F^{\alpha \beta}(\theta,\eta) = F^{\alpha \beta}_{\text{free}}(\theta, \eta)
\underbrace{P(\cosh(\theta -\eta))F_{\text{min}}(\theta -\eta +i\pi)}_{=:F_P(\theta-\eta)}
\widetilde{g^2} (\mu\cosh\theta-\mu\cosh\eta),
\end{equation}
where $\mu>0$ is the mass of the particle and 
\begin{equation}
F^{\alpha \beta}_{\text{free}}(\theta,\eta) = \frac{\mu^2}{2\pi}
\left( \begin{array}{ccc}
\cosh^2\big( \frac{\theta +\eta}{2}\big) & \frac{1}{2}\sinh (\theta +\eta ) \\
\frac{1}{2}\sinh(\theta +\eta) & \sinh^2 \big( \frac{\theta +\eta}{2} \big)  \end{array} \right).
\end{equation}
Here $F^{\alpha \beta}_{\text{free}}$ is the well-known expression of the ``canonical'' stress-energy tensor of the free Bose field, and $F_{\text{min}}$ is the so called \emph{minimal solution} of the model\cite{KarowskiWeisz:1978}, which is unique for a given scattering function. For example, $F_{\text{min}}(\zeta)=1$  for free fields and $F_{\text{min}}(\zeta)=-i\sinh \frac{\zeta}{2}$ in the Ising model; for the sinh-Gordon model, $F_{\text{min}}$ is given as an integral expression\cite{FringMussardoSimonetti:1993}.

\section{Negative energy density and QEIs in one-particle states }

First let us investigate whether there are single-particle states with negative energy density.

In the case of free fields (with $P=1$), it is known that there are no such states: One has to allow for superpositions, for example, of a zero- and a two-particle vectors to obtain negative energy densities. 

However, the introduction of interaction changes this situation drastically. Specifically, if there is a $\theta_P \in \mathbb{R}$ such that $|F_P(\theta_P)|>1$, then there exists a one-particle state $\varphi \in L^2(\mathbb{R})$ and a real-valued Schwartz function $g$ such that $\langle \varphi, T^{00}(g^2) \varphi \rangle < 0$ (see Prop. 4.1 in Ref.~\citenum{BostelmannCadamuro:oneparticle}.)
This applies, in particular, to the Ising and sinh-Gordon models.

Under stronger assumptions on the function $F_P$ we can actually prove a \emph{no-go} theorem on existence of QEIs at one-particle level (see Ref.~\citenum{BostelmannCadamuro:oneparticle}, Proposition~4.2):
\begin{theorem}\label{nogo}
Suppose there exist $\theta_0 \geq 0$ and $c > \frac{1}{2}$ such that
\begin{equation}
\forall \theta \geq \theta_0: \quad F_P(\theta) \geq c \cosh\theta.
\end{equation}
Let $g$ be real-valued and of Schwartz class, $g \not\equiv 0$. Then there exists a sequence $(\varphi_j)_{j \in \mathbb{N}}$ in $\mathcal{D}(\mathbb{R})$, $\| \varphi_j \| =1$, such that
\begin{equation}
\langle \varphi_j, T^{00}(g^2) \varphi_j \rangle \rightarrow -\infty \quad \text{as } j \rightarrow \infty.
\end{equation}
\end{theorem}
Here $\mathcal{D}(\mathbb{R})$ denotes the space of $\mathcal{C}^\infty$-functions with compact support. This proposition says that if the function $F_P$  grows ``too fast'', then the operator $T^{00}(g^2)$ (at one-particle level) cannot be bounded below. Only under certain upper bounds on $F_P$, we can establish one-particle state-independent QEIs (Thm.~5.1 of Ref.~\citenum{BostelmannCadamuro:oneparticle}):
\begin{theorem}\label{theoqei}
Suppose there exist $\theta_0 \geq 0, \lambda_0 >0$, and $0<c< \frac{1}{2}$ such that
\begin{equation}\label{boundfp}
|F_P(\zeta)| \leq c \cosh \operatorname{Re} \zeta \quad \text{whenever } \lvert\operatorname{Re} \zeta\rvert \geq \theta_0, \; 
\lvert\operatorname{Im} \zeta\rvert< \lambda_0.
\end{equation}
Let $g$ be a real-valued Schwartz function, then there exists $c_g >0$ such that
\begin{equation}\label{oneqei}
\forall \varphi \in \mathcal{D}(\mathbb{R}): \quad \langle \varphi, T^{00}(g^2) \varphi \rangle \geq - c_g \| \varphi \|^2.
\end{equation}
The constant $c_g$ depends on $g$ (and on $F_P$, hence on $P$ and $S$) but not on $\varphi$.
\end{theorem}
To motivate the above condition on the growth property of $F_P$, let us consider the expectation value of $T^{00}(g^2)$ in a one-particle state $\varphi$,
\begin{equation}\label{expectation}
\langle \varphi, T^{00}(g^2) \varphi \rangle = \frac{\mu^2}{2\pi} \int d\theta d\eta\; \cosh^2\frac{\theta +\eta}{2} F_P(\theta -\eta)\widetilde{g^2}(\omega(\theta) -\omega(\eta)) \overline{\varphi(\theta)}\varphi(\eta),
\end{equation}
where $\omega(\theta) := \mu \cosh \theta$. 

To establish an inequality of the form Eq.~\eqref{oneqei}, the rough idea goes as follows. The relevant contributions to the integral are in the regions $\theta \approx \eta$ and $\theta \approx -\eta$; outside these regions, the factor $\widetilde{g^2}(\omega(\theta) -\omega(\eta))$ is strongly damping. At $\theta \approx \eta$, the factor $F_P$ is nearly constant, and hence the integral is near to the free field expression, which is known to be positive when the smearing function is of the form $g^2$. At $\theta \approx -\eta$, the factor $F_P$ may grow but the factor $\cosh \frac{\theta +\eta}{2}$ is nearly constant. If now $F_{\text{min}}$ grows less than $\frac{1}{2}\cosh\theta$, then the second mentioned part of the integral is negligible against the first mentioned one and the whole expression is positive, up to some bounded part. 

In other words, at one-particle level the existence or non-existence of QEIs depends on the asymptotic behaviour of the function $F_P(\theta) = P(\cosh\theta)F_{\text{min}}(\theta +i\pi)$. If $F_P(\theta) \lesssim \frac{1}{2}\cosh\theta$, then a state-independent one-particle QEI holds. On the other hand, if $F_P(\theta) \gtrsim \frac{1}{2}\cosh\theta$, then no such QEI holds (see Proposition~\ref{nogo}).

In specific models, the bound \eqref{boundfp} is fulfilled for at least some choices of $P$: In the free and sinh-Gordon models, the function $F_{\text{min}}$ converges to a constant for large $\theta$, thus a QEI can hold only if $\text{deg }P=0,1$.

In the Ising model, where the function $F_{\text{min}}(\theta +i\pi)$ grows like $\cosh \frac{\theta}{2}$ at large values of $\theta$, a QEI holds if and only if $P \equiv 1$.

In other words, for some $S$ the existence of QEIs fixes the energy density uniquely at one-particle level; whereas, in the sinh-Gordon model, and even in the free Bose field, we are left with the choice
\begin{equation}
P(x)=(1-\alpha) +\alpha x \quad \text{with } \alpha \in \mathbb{R},\;  |\alpha|< \frac{1}{2F_{\text{min}}(\infty +i\pi)},
\end{equation}
where $F_{\text{min}}(\infty +i\pi) := \lim_{\theta \rightarrow \infty}F_{\text{min}}(\theta +i\pi)$. Therefore, in this second class of models the existence of a QEI strongly restricts the form of the energy density without, however, fixing it uniquely.

\section{Numerical results}

While we now have upper and lower estimates for the lowest eigenvalue in the spectrum of $T^{00}(g^2)$ on the one-particle level, it seems unrealistic to find the actual lowest eigenvalue in this way to any reasonable precision. An explicit result can be obtained only numerically. Here we sketch some results of Ref.~\citenum{BostelmannCadamuro:oneparticle}.

For the sake of concreteness, we fix the smearing function $g$ to be a Gaussian,
\begin{equation}
g(t) = \pi^{-1/4}\sqrt{\frac{\mu}{2\sigma}} \exp\Big( -\frac{(\mu t)^2}{8\sigma^2}\Big),
\end{equation}
where $\sigma>0$ is a dimensionless parameter. 

For the numerical treatment, we restric the one-particle wavefunctions of the matrix elements of $T^{00}(g^2)$ to the Hilbert space $L^2([-R,R],d\theta)$ rather than $L^2(\mathbb{R},d\theta)$, that is, we introduce a ``rapidity cutoff''. This serves to make the kernel yield a bounded operator. We then discretize the operator by dividing the interval $[-R,R]$ into N subintervals and using an orthonormal system of step functions $\phi_j$ supported on these intervals. We are then left with a matrix
\(
M_{jk}= \langle \phi_j, T^{00}(g^2) \phi_k \rangle
\)
for which eigenvalues and eigenvectors can be found by standard numerical methods, such as the implicit QL algorithm.

The eigenvector corresponding to the lowest eigenvalue for the sinh-Gordon model is shown in \fref{fig:sinh}(a). We can then analyze how the lowest eigenvalue (i.e., the best constant $c_g$ in the QEI) depends on the interaction. \fref{fig:sinh}(b) shows the dependency on the coupling constant $B$ in the sinh-Gordon model. As expected, as the coupling is taken to 0, we reach the limit 0 for the lowest eigenvalue (as in free field theory). Moreover, we note that the lowest possible negative energy density is reached when the coupling is maximal $(B=1)$, which fits with the picture that negative energy density in one-particle states is an effect of self-interaction in the quantum field theory.

\def\figsubcap#1{\par\noindent\centering\footnotesize(#1)}
\begin{figure}[th]%
\begin{center}
  \parbox{2.1in}{\includegraphics[width=2in]{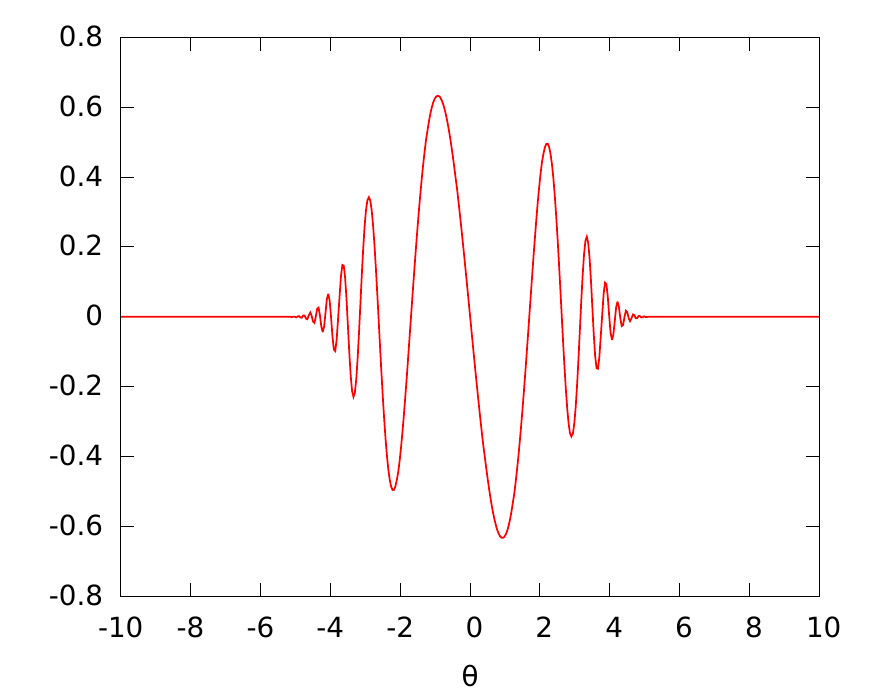}\figsubcap{a}}
  \hspace*{4pt}
  \parbox{2.1in}{\includegraphics[width=2in]{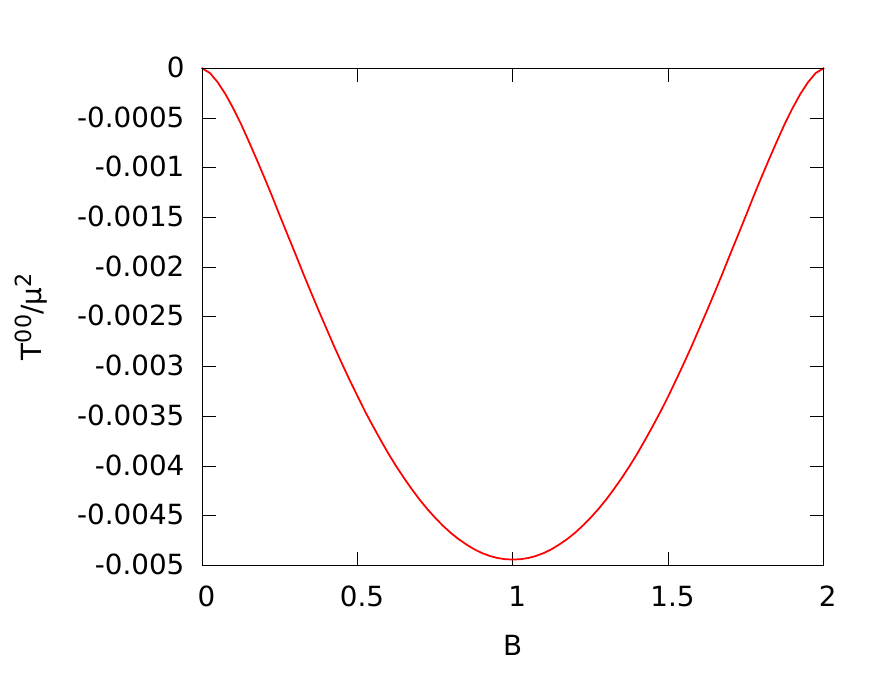}\figsubcap{b}}
  \caption{Energy density $T^{00}(g^2)$ in the sinh-Gordon model at one-particle level. (a) Lowest eigenvector for $B=1$. (b) Lowest eigenvalue for coupling $0 < B< 2$. Parameters: $N=500, R= (a)10\; (b)7, \sigma=0.1, P=1$.}%
  \label{fig:sinh}
\end{center}
\end{figure}

\section{Conclusions}

We have investigated the properties of the energy density at one-particle level in a large class of quantum integrable models, which include the sinh-Gordon and Ising models. In particular, we have determined the stress-energy tensor at one-particle level up to a certain polynomial factor, and studied the existence of state-independent QEIs. Moreover, we have seen that demanding the existence of QEIs can, in some cases, fix this choice uniquely.

An investigation of analogous results at higher particle numbers is a challenging open question (except for the Ising model where results are available in Ref.~\citenum{BostelmannCadamuroFewster:ising}) since higher form factors $F_{n}$ have poles on the integration path, and the integral operator $T^{00}(g^2)$ is hard to estimate (singular integral kernels). Numerical evidence, however, suggest that a term like \eqref{expectation} is the dominating contribution also at two-particle level.

Apart from scalar models, existence of QEIs could be investigated also in models with more then one particle species (e.g. the nonlinear $O(N)$--invariant $\sigma$--models), or in integrable models with bound states ($Z(N)$--Ising and sine--Gordon models), and it would be highly desirable to extend this study to a curved background. This would be a possible path towards a model-independent understanding of the energy density and QEIs.

\bibliographystyle{ws-procs975x65}
\bibliography{integrable}

\begin{thebibliography}{10}

\bibitem{HawPen1970}
S.~W. Hawking and R.~Penrose, The singularities of gravitational collapse and
  cosmology, {\em Proc. Roy. Soc. London Ser. A} {\bf 314}, 529  (1970).

\bibitem{Hawking:1992}
S.~W. Hawking, Chronology protection conjecture, {\em Phys. Rev. D} {\bf 46},
  603  (1992).

\bibitem{Ford:1991}
L.~H. Ford, Constraints on negative-energy fluxes, {\em Phys. Rev. D} {\bf 43},
  3972  (1991).

\bibitem{Fewster:lecturenotes}
C.~J. Fewster, Lectures on quantum energy inequalities arXiv:1208.5399,
  (2012).

\bibitem{BostelmannFewster:2009}
H.~Bostelmann and C.~J. Fewster, Quantum inequalities from operator product
  expansions, {\em Commun. Math. Phys.} {\bf 292}, 761  (2009).

\bibitem{BostelmannCadamuroFewster:ising}
H.~Bostelmann, D.~Cadamuro and C.~J. Fewster, Quantum energy inequality for the
  massive {Ising} model, {\em Phys. Rev. D} {\bf 88}, p. 025019 (Jul 2013).

\bibitem{BabujianFoersterKarowski:2006}
H.~M. Babujian, A.~Foerster and M.~Karowski, The form factor program: {A}
  review and new results, {\em SIGMA} {\bf 2}, p. 082  (2006).

\bibitem{Lechner:2008}
G.~Lechner, Construction of quantum field theories with factorizing
  {S}-matrices, {\em Commun. Math. Phys.} {\bf 277}, 821  (2008).

\bibitem{FringMussardoSimonetti:1993}
A.~Fring, G.~Mussardo and P.~Simonetti, Form-factors for integrable
  {L}agrangian field theories, the sinh-{G}ordon model, {\em Nucl. Phys.} {\bf
  B393}, 413  (1993).

\bibitem{BostelmannCadamuro:oneparticle}
H.~Bostelmann and D.~Cadamuro, Negative energy densities in integrable quantum
  field theories at one-particle level arXiv:1502.01714.

\bibitem{BostelmannCadamuro:characterization}
H.~Bostelmann and D.~Cadamuro, Characterization of local observables in
  integrable quantum field theories, {\em Commun. Math. Phys.} {\bf 337}, 1199
  (2015).

\bibitem{KarowskiWeisz:1978}
M.~Karowski and P.~Weisz, Exact form factors in (1 + 1)-dimensional field
  theoretic models with soliton behaviour, {\em Nucl. Phys.} {\bf B139}, 455
  (1978).

\end{thebibliography}

\end{document}